\documentclass[aps,twocolumn,preprintnumbers,amsmath,amssymb,superscriptaddress]{revtex4-1}%
\usepackage{amsthm,bbm}
\usepackage{graphicx,epsfig,epsf,color,hhline}
\usepackage{dcolumn}
\usepackage{bm}
\usepackage{dsfont}
\usepackage[english]{babel}
\usepackage{amsmath}
\usepackage{amsfonts}
\usepackage{amssymb}
\usepackage{graphicx}

\begin{document}

\title{Discord, quantum knowledge and private communications}
\author{Mile Gu}
\affiliation{School of Physical and Mathematical Sciences, Nanyang
Technological University, Singapore 639673}
\affiliation{Complexity Institute, Nanyang Technological
University, Singapore 637723}
\affiliation{Centre~for~Quantum~Technologies,
National~University~of~Singapore, 117543, Singapore}
\author{Stefano Pirandola}
\affiliation{Computer Science \& York Centre for Quantum
Technologies, University of York, York YO10 5GH, UK}

\begin{abstract}
In this brief review, we discuss the role that quantum
correlations, as quantified by quantum discord, play in two
interesting settings. The first one is discerning which unitaries
have been applied on a quantum system, by taking advantage of
knowledge regarding its initial configuration. Here discord
captures the `quantum' component of this knowledge, useful only
when we have access to a quantum memory. In particular, discord
can be used to detect whether an untrusted party has certain
quantum capabilities. The second setting is quantum cryptography.
Here discord represents an important resource for trusted-noise
quantum key distribution and also provides a general upper bound
for the optimal secret key rates that are achievable by ideal
protocols. In particular, the (two-way assisted) secret key
capacity of a lossy bosonic channel exactly coincides with the
maximum discord that can be distributed between the remote parties
at the two ends of the channel.
\end{abstract}

%\affiliation{Complexity Institute, Nanyang Technological
%University, 60 Nanyang View, Singapore 639673, Republic of
%Singapore} \affiliation{Centre for Quantum Technologies, National
%University of Singapore,Republic of Singapore}

\maketitle

\section{Knowledge, Correlations, and Guessing Channels}

The 1962 James Bond's movie `Dr. No' taught children around the world a
valuable lesson in how to detect whether nosy siblings are snooping into their
rooms. You stick a small piece hair across the door and the doorframe. When
the door is opens, the hair falls to the floor. The unsuspecting perpetrator
has unwittingly communicated to you their rather unscrupulous action. This
trick demonstrates the power of knowledge; by knowing how a system is
initially configured (the location of hair), one can gain information about
actions that have affected the system (opening the door).

This phenomena can be described by information theory. We denote a system of
interest to be $A$, and knowledge about the system to be encoded within some
memory $B$ - an approach previously adopted to understand uncertainty
relations under quantum memory~\cite{berta10}. If $B$ contains information
about $A$, the two systems will be correlated, such that $I(A,B) > 0$.

The classical one time pad provides a simple example. Here Alice and Bob wish
to communicate some secret message in the future. To do this, Alice and Bob
gather in some secure location, where Alice generates a string of random bits
that Bob commits them to memory. That is, they share many copies of the
classically correlated state
\begin{equation}
\rho= |00\rangle\langle00| + |11\rangle\langle11|.
\end{equation}
Should Alice choose to flip some of her bits and give the resulting string to
Bob, Bob is able to discern exactly which bits have been flipped by comparing
the resulting string with the one stored in his memory. In contrast, anyone
without access to Bob's memory would gain no information about Alice's
actions. The optimal such scheme would allow Alice to communicate $\delta_{I}
= 1$ bit per copy of $\rho_{AB}$ shared. Thus possession of $B$ allows
exclusive knowledge of how the system was manipulated. One notes that here,
$I(A,B) = 1$, which is equal to $\delta_{I}$. This is in fact, not a coincidence.

Consider the following general \textquotedblleft channel guessing
game\textquotedblright.

\begin{enumerate}
\item Alice and Bob initially share a state $\rho$ distributed over the system
of interest $A$, and the memory $B$. This initial state is publicly known.

\item Alice applies some unitary operator $U_{k}$ onto her subsystem $A$ with
probability $p_{k}$. She publicly announces her protocol (e.g. the unitaries
$U_{k}$ and their probability of application), but not the specific $k$ she
selects in each run.

\item Alice gives $A$ to Bob, so that Bob is now in possession of $\rho
_{AB}^{(k)}=U_{k}\rho_{AB}U_{k}^{\dag}$. Without knowledge of $k$, Bob sees
the ensemble state $\tilde{\rho}_{AB}=\sum_{k}p_{k}\rho_{AB}^{(k)}$.

\item Alice challenges Bob to guess which $U_{k}$ she has applied, i.e., to
estimate the value of $k$.
\end{enumerate}

This game captures a communication channel between Alice and Bob, where Alice
has encoded a random variable $K$ that takes the value $k$ with probability
$p_{k}$, onto corresponding codewords $\rho_{AB}^{(k)}$. The maximum
information rate of this channel is then bounded above by the Holevo quantity
\begin{equation}
I_{q}=\tilde{S}(A,B)-S(A,B), \label{eq:iq}%
\end{equation}
where $S(A,B)$ and $\tilde{S}(A,B)$ represent the respective entropies of
$\rho_{AB}$ and $\tilde{\rho}_{AB}$. Here we consider the i.i.d. limit of many
trials, where Alice repeats this game a large number of times; the performance
of Bob, as quantified by the maximum information per trial, then saturates
$I_{q}$.

This relation has a nice interpretation. In fact $\tilde{\rho}$ describes the
state of the bipartite system after encoding, as viewed by an observer who is
unaware of which $k$ was encoded in each run. Therefore $\tilde{S}%
(A,B)-S(A,B)$ captures the gain in entropy (or alternatively, the cost in
negentropy) of encoding $K$ from their perspective. Thus Eq.~(\ref{eq:iq})
tells us that communication of $k$ bits of data necessarily incurs a minimum
entropic cost of $k$.

Suppose Bob cannot access his memory (e.g. it was lost), the effective
codewords would now be be $\rho_{A}^{(k)}$, with associated Holevo quantity
\begin{equation}
I_{0}=\tilde{S}(A)-S(A).
\end{equation}
The impact of having memory on Bob's in performance at the i.i.d. limit is
then
\begin{equation}
\Delta_{q}\equiv I_{q}-I_{0}=I(A,B)-\tilde{I}(A,B),
\end{equation}
where $I(A,B)$ and $\tilde{I}(A,B)$ are the respective mutual information of
$\rho_{AB}$ and $\tilde{\rho}_{AB}$. The quantity $I(A,B)-\tilde{I}(A,B)$ then
represents the cost, in terms of total correlations between $A$ and $B$ of
encoding $K$. Meanwhile $\Delta_{q}$ represents information about $K$ that is
exclusively available to Bob due to his possession of $B$.

If we consider $I(A,B)$ to capture the amount of knowledge Bob knows about
$A$, we find an interesting resource based view of knowledge: \emph{Bob can
expend $k$ bits of knowledge about a system $A$ to learn at most $k$ bits of
information about actions on $A$; in the i.i.d limit, this bound can be
saturated}. That is, knowing $k$ bits about some system $A$, as captured by
possessing a system $B$ such that $I(A,B)=k$, implies that one can gain up to
$k$ extra bits about actions on $A$. Thus for the one-time pad, a shared
mutual information of $1$ allows Alice to securely communicate a single bit to
Bob. Meanwhile in quantum dense coding, Alice and Bob initially share a Bell
state - such that $I(A,B)=2$. Thus, Bob can harness his memory to gain $2$
exclusive bits about Alice's actions on $A$.

\subsection{\textbf{The Role of Discord}}

Recall that we can separate correlations into two components, i.e., we can
write $I(A,B)=J(A|B)+\delta(A|B)$, where $J(A|B)$ and $\delta(A|B)$
respectively represent the purely-classical correlations and the quantum
correlations (discord). This fits well with the channel guessing game. The
approach is to relate quantum and classical correlations with the quantum and
classicality of Bob's memory. Specifically let us consider the three scenarios:

\begin{enumerate}
\item \emph{Memoryless Bob:} Bob's memory is completely faulty. That is, Bob
cannot access $B$ at all. Bob's resulting performance is then given by $I_{0}$
(as defined above).

\item \emph{Classical Bob:} Bob's memory is classical. That is, Bob is
required to measure any $\rho_{b}$ given to him with respect to some
orthogonal basis, and stored the measurement results in place of $\rho_{b}$.
Denote Bob's resulting performance by $I_{c}$.

\item \emph{Quantum Bob:} Bob has unrestricted quantum information processing,
and can (i) store $\rho_{b}$ without error, and (ii) coherently interact his
memory with the system of interest. Bob's resulting performance is given by
$I_{q}$.
\end{enumerate}

\begin{figure*}[ptb]
\includegraphics[width=0.95\textwidth]{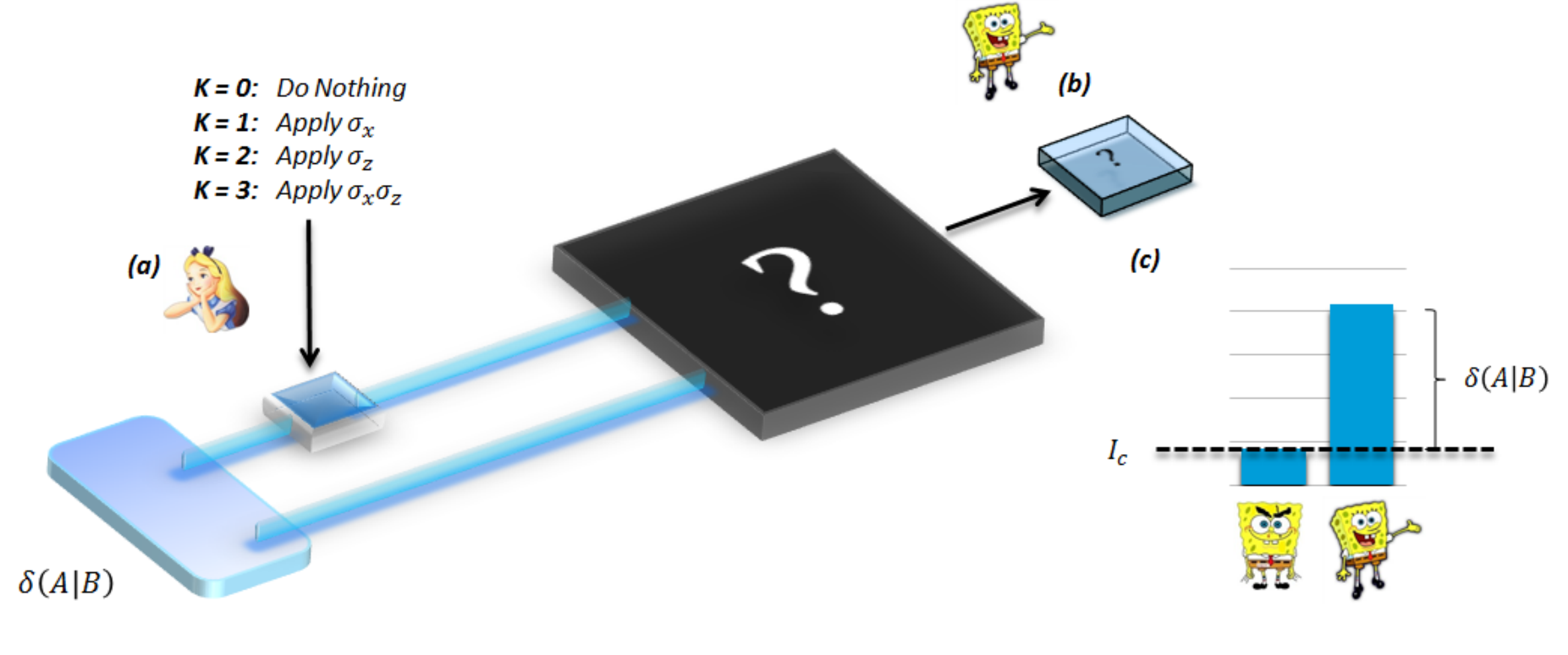}
\caption{\textbf{Example on
Two Qubits}. Consider the special case where Alice and Bob share a correlated
state on two qubits, $A$ and $B$, with discord $\delta(A|B)$. Alice then
encodes a random variable $K$ governed by a uniform distribution over
$\{0,1,2,3\}$ by applying one of four possible unitaries, $I$, $\sigma_{x}$,
$\sigma_{z}$ or $\sigma_{x}\sigma_{z}$ and challenges Bob to estimate $K$. In
this scenario, the encoding is maximal, and Bob's performance gain when using
quantum in the place of classical memory is given exactly by $\delta(A|B)$.
This protocol has been experimentally implemented by by Almeida
et.al.~\cite{almeida12}.}%
\label{fig:test}%
\end{figure*}

Cases 1 and 3 have been outlined above. Our focus here is thus case $2$. The
rationale is that a classical memory should be able to make use of purely
classical correlations, but not quantum correlations. Therefore, we would
expect discord to be related with the performance gap, $I_{q}-I_{c}$, between
quantum and classical Bob. This problem was studied in Gu. et.
al.~\cite{mgu12}, where they established that
\begin{equation}
J(A|B)-\tilde{J}(A|B)\leq I_{c}-I_{0}\leq J(A|B).
\end{equation}
Here $\tilde{J}(A|B)$ and $\tilde{J}(A|B)$ represent the classical
correlations in $\rho_{AB}$ and $\tilde{\rho}_{AB}$. The equation describes an
interesting connection between discord and the performance advantage of having
quantum-over-classical memory. That is, introducing $\delta(A|B)-\tilde
{\delta}(A|B)$ as the discord difference before and after encoding, we get
\begin{equation}
\Delta{\delta}(A|B)-\tilde{I}(A,B)\leq I_{q}-I_{c}\leq\Delta{\delta}(A|B),
\end{equation}
where $\tilde{I}(A|B)$ is the mutual information of $\tilde{\rho}$. Consider
now any encoding that attempts to communicate the maximum amount of
information (known as a maximal encoding). In this scenario, $\tilde{\rho}%
_{A}$ is maximally mixed, $\tilde{\delta}=\tilde{I}=0$, and thus we have:

\begin{enumerate}
\item $I_{0} = 1 - S(A)$: a memoryless Bob can only access the local memory
available on $A$. That is, the maximum amount of information Bob can learn
about what happens to $A$ is exactly the negentropy of $A$.

\item $I_{c} = I_{0} + J(A|B)$: a classical Bob can learn an additional
$\Delta_{c} = J(A|B)$ bits of information about actions on $A$. That is, he
can exactly take advantage of the classical correlations between $A$ and $B$.

\item $I_{q} = I_{0} + I(A,B)$: a quantum Bob can take advantage of the full
correlations between $A$ and $B$. As such, his performance advantage over the
classical case is exactly $\delta(A|B)$, the discord between $B$ and $A$.
\end{enumerate}

These relations capture an operational interpretation of discord $\delta(A|B)$
as how much purely quantum mechanical knowledge $B$ has about $A$. An example
if given in Fig. \ref{fig:test}.

\subsection{\textbf{Example: Certifying Entangling Gates without
Entanglement}}

The interpretation of discord as quantum knowledge can be applied to verify
whether someone is in possession of entangling gates, as also experimentally
realized by using polarization photons~\cite{almeida12}. Consider the case
where Bob claims that he is capable of building entangling two-qubit gates.
How can Alice verify that Bob is telling the truth - without being able to
generate entanglement herself?

The inability for classical processors to harness quantum knowledge suggests
an immediate solution. Suppose now Alice prepares some discord, two-qubit
state, $\rho_{AB}$. She can then perform the protocol above, using a specific
encoding scheme that encodes two bits, $a, b \in\{0,1\}$, onto $A$, by
applying the unitary $U = X^{a}Z^{b}$, where $X$ and $Z$ are standard Pauli
operators. This corresponds to a scenario where $\tilde{\rho}_{A} = I/2$ is
maximally mixed. Alice then challenges Bob guess $a$ and $b$. Bob's
performance is then characterized by the mutual information between the
encoded bits, and that of Bob's guess.

In the 2 qubit case, it can be shown that if Bob is incapable of synthesizing
entangling two-qubit gates, then he cannot exceed the performance level of
$I_{c}$. As such any performance exceeding $I_{c}$ implies that Bob is capable
of some entangling operations. Thus, discord can be used as a way of
certifying entanglement without entangling gates.

\section{Discord in quantum key distribution}

Quantum discord\ also plays an important role in private communications and
quantum key distribution (QKD)~\cite{Gisin02,ScaraniRMP,WeeRMP}. The fact that
it must be non-zero is intuitive: Quantum discord and its geometric
formulation are connected with the concept of non-orthogonality, which is the
essential ingredient for QKD. A scenario where this is particularly evident is
device-dependent (or trusted-device) QKD. This includes all those realistic
situations where the noise affecting the local devices is assumed to be
trusted. For instance this can be detection noise (genuine inefficiency or
noise added by the parties~\cite{Renner2005,Pirandola09}) or preparation
noise, as in the settings of untrusted-relay
QKD~\cite{Braunstein12,Pirandola15b} and
thermal-QKD~\cite{Filip08,Usenko10,Weedbrook2010,Weedbrook2012,Weedbrook2014}.
Such trusted noise may be so high to prevent any entanglement distribution,
but still a secure key\ can be extracted due to non-zero discord.

Any QKD\ protocol can be recast into a measurement-based scheme, where Alice
sends Bob part of a bipartite state, then subject to local detections. Let us
describe a device-dependent protocol in this representation. In her private
space, Alice prepares two systems, $A$ and $a$, in a generally mixed state
$\rho_{Aa}$. This state is purified into a 3-partite state $\Phi_{PAa}$ with
the ancillary system $P$ being inaccessible to Alice, Bob or Eve. This system
accounts for the trusted noise in Alice's side. Then, system $b$ is sent to
Bob, who gets the output $B$ after the channel (eavesdropping). Bob's output
$B$ is assumed to be affected by other local trusted noise in Bob's private
space (denoted as $P$ as before). Finally, from the shared state $\rho_{AB}$,
Alice and Bob extract two correlated variables by applying suitable
measurements. On the output data, they perform error correction and privacy
amplification with the help of one-way CC, which can be either forward (direct
reconciliation, $\blacktriangleright$), or backward (reverse reconciliation,
$\blacktriangleleft$).

They finally extract a key at a rate $R=\max\{R_{\blacktriangleright
},R_{\blacktriangleleft}\}$, maximised between the reconciliations. Now we
have~\cite{Pirandola14}
\begin{equation}
E_{D}(A,B)\leq R\leq E_{D}(A,B)+I(AB,P)~, \label{QKDdiscord}%
\end{equation}
where $E_{D}(A,B)$ is the one-way distillable entanglement for systems $A$ and
$B$, as quantified by the maximum between the
coherent~\cite{Schumacher96,Lloyd97} and reverse coherent
information~\cite{Patron09,Pirandola2009}, while $I(AB,P)$ is the quantum
mutual information between $AB$ and the trusted-noise system $P$. From
Eq.~(\ref{QKDdiscord}), we see that the existence of $P$ is necessary in order
to have $R>0$ in the absence entanglement (i.e., for $E_{D}=0$). Indeed it is
easy to find discord-based Gaussian QKD protocols for which this is
possible~\cite{Pirandola14}. According to Eq.~(\ref{QKDdiscord}), the absence
of $P$ implies $R=E_{D}$, so that secure key distribution becomes equivalent
to entanglement distillation~\cite{Ekert91}.

In the absence of trusted noise, we have ideal QKD protocols where all the
noise in the global output state is partly controlled by the parties and
partly by Eve. In this setting, quantum discord becomes a simple upper bound
for the key rate. In fact, for any ideal QKD\ protocol in direct or reverse
reconciliation, we may write~\cite{Pirandola14}
\begin{equation}
R\leq\max\{\delta(A|B),\delta(B|A)\},
\end{equation}
where $\delta(A|B)$ and $\delta(B|A)$ are the two types of discord.
Surprisingly, for the important practical case of a lossy
channel~\cite{WeeRMP} with transmissivity $\eta$, such as an optical fiber or
a free-space link, the previous bound becomes tight. This is due to a
combination of elements. First of all, we may always write~\cite{Pirandola14}
\begin{equation}
R_{\blacktriangleleft}=\delta(B|A)-E_{F}(B,E),
\end{equation}
where $E_{F}(B,E)$ is the entanglement of formation between Bob and Eve.
Second, the Stinespring dilation of a lossy channel is a beam splitter with
transmissivity $\eta$, mixing the Alice's input state with a vacuum
environmental mode. For this reason, Bob and Eve's output state is not
entangled, i.e., $E_{F}(B,E)=0$. Therefore, in a lossy channel, we always
have
\begin{equation}
R_{\blacktriangleleft}=\delta(B|A).
\end{equation}

Most importantly, one can prove~\cite{Pirandola15} that the maximum discord
$\delta_{\max}(B|A)$ that can be distributed to the parties through the lossy
channel coincides with the secret-key capacity $K$ of the lossy channel (where
this capacity is generally defined assuming the most general feedback-assisted
protocols for key generation, based on unlimited two-way CC and adaptive local
operations). In fact, Ref.~\cite{Pirandola15} showed that%
\begin{equation}
K(\eta)=\delta_{\max}(B|A)=-\log_{2}(1-\eta)~,\label{PLOB}%
\end{equation}
which provides the ultimate rate-loss scaling for bosonic secure
communications, approximately $1.44\eta$ secret bits per channel use for high
loss (i.e., at long distances)

The proof Eq.~(\ref{PLOB}) is based on several ingredients. First
of all, it exploits the technique of teleportation stretching,
devised in Ref.~\cite{Pirandola15} for point-to-point
quantum/private communications, and then extended in
Ref.~\cite{Network} to quantum repeaters and communication
networks, and in Ref.~\cite{CosmoPir} to quantum metrology and
channel discrimination. In this technique, an arbitrary adaptive
protocol for quantum/private communication is simplified into a
much simpler non-adaptive form, providing the same output state as
the original one. The advantage is
that such output state is now decomposed in the form $\bar{\Lambda}%
(\rho_{\mathcal{E}}^{\otimes n})$, where $\bar{\Lambda}$ is a trace-preserving
LOCC, $\rho_{\mathcal{E}}$ is the Choi matrix~\cite{Choi} of the channel
$\mathcal{E}$ (to be defined as suitable limit for a lossy channel), and $n$
is the number of uses of the channel. This decomposition is possible because
the lossy channel is covariant with respect to the displacement operators and
therefore can be simulated by means of continuous variable quantum
teleportation~\cite{SamTele,reviewtele}. In other words, the lossy channel is
a specific example of teleportation-covariant channel~\cite{Pirandola15}.

The second ingredient is introduction of the channel's relative entropy of
entanglement $E_{R}(\mathcal{E})$, which extends the original definition for
quantum states~\cite{REE1,REE2,REE3} to quantum channels.
Ref.~\cite{Pirandola15}\ proved that, for any channel $\mathcal{E}$, the
secret-key capacity satisfies the bound $K(\mathcal{E})\leq E_{R}%
(\mathcal{E})$ (see also Ref.~\cite{Ric17}). For the specific case
of the lossy channel, one may combine
the Choi-decomposition of the output $\bar{\Lambda}(\rho_{\mathcal{E}%
}^{\otimes n})$ together with the properties of the relative entropy of
entanglement to prove that $K(\eta)\leq E_{R}(\rho_{\mathcal{E}})$. The latter
term is the relative entropy of entanglement of the asymptotic Choi matrix of
the lossy channel and must be computed as a limit over a sequence of two-mode
squeezed vacuum states~\cite{Pirandola15}. This procedure leads to the upper
bound
\begin{equation}
K(\eta)\leq-\log_{2}(1-\eta).
\end{equation}
Since the upper bound is achievable by a suitable Gaussian protocol in reverse
reconciliation~\cite{Pirandola2009,Pirandola14}, we then achieve
Eq.~(\ref{PLOB}). The proof can be easily extended to include the two-way
quantum capacity, so that we also have $K(\eta)=Q_{2}(\eta)$%
~\cite{Pirandola15}.

\section{Conclusions}

In this brief review, we have discussed the role that quantum discord plays in
two interesting settings. First of all, we considered the scenario of a
bipartite system consisting of a system of interest, $A$, and a memory system
$B$, such that their correlations, $I(A,B)$, represent knowledge $B$ has about
$A$. This knowledge can be harnessed by a person in possession of $B$ to gain
extra information about what performed on $A$. In this context, we outlined
how discord is captured in the quantum component of such knowledge - measuring
the component of $I(A,B)$ that is useful only when $B$ can be stored in
quantum memory.

We then reviewed how quantum discord can be seen as a primitive for quantum
cryptography, where it plays a double role. It is the bipartite resource which
is exploited in trusted-noise QKD, where the presence of such noise may
prevent the exploitation of quantum entanglement but not the distribution of a
secret key. Then, quantum discord provides a general upper bound to the key
rate in the ideal case when trusted noise is absent. In particular, this bound
is achievable in the important case of lossy bosonic communications. In this
setting, the maximum discord that two remote parties can generate at the two
ends of a lossy channel corresponds exactly to the maximum number of secret
bits that they can generate through the channel by means of the most general
adaptive protocols for QKD.

\end{document}